# A Distributed Step-by-step Finite-time Consensus Design for Heterogeneous Battery Energy Storage Devices with Droop Control

Yalin Zhang, Yunzhong Song and Shumin Fei

***Abstract*— As all generators are distributed in different areas among large scale power systems, the cooperative manipulation of the multi-generator system cannot be done well without consideration of the distance information of the generators, A distributed step-by-step finite-time consensus scheme for the heterogeneous Battery Energy Storage System (BESS) is proposed in this paper, where the coordinated consensus can be come into reality within a limited time, which is appealing for the electrical engineering community. To be concrete, at first, all BESSs are classified into several clusters according to their locations, and in each cluster, there is an active leader in charge of information receiving from outside. Then after, in order to coordinate the multi BESSs, five inputs, which are function oriented, were used to achieve energy level balancing, active/reactive power sharing, and voltage/frequency synchronization of the multi BESSs. To be further, the frequency and voltage restoration to the nominal values of the main grid was made possible by the introduction of a virtual leader, which is actually an external leader. Compared with the centralized methods, this control scheme is entirely distributed, and each BESS only utilizes the information of its own and its neighbors. Besides, this control is robust to the load perturbation and the plug-and-play of the communication topology. Finally, some simulation experiments are executed on the modified IEEE 57-bus system to verify the suggested scheme.**

***Index Terms*—Battery energy storage system, finite-time consensus, multi-agent system, distributed cooperative control, step-by-step consensus.**

## NOMENCLATURE

| | |
|---|---|
| BESS | Battery Energy Storage System |
| $i$ , $j$ | The index of agents |
| $v_{odi}$ , $v_{oqi}$ , $i_{odi}$ and $i_{oqi}$ | The d-axis and q-axis of voltage and current off BESS $i$ |
| $P_i$ , $Q_i$ | The calculated value of BESS $i$ |
| $\omega_{ci}$ | The cutoff frequency of the low-pass filter of BESS $i$ |
| $K_i^P$ , $K_i^Q$ | The droop coefficients of BESS $i$ |
| $\omega_i$ , $V_i$ , $P_i$ , $Q_i$ and $E_i$ | The frequency, voltage, active/reactive power and energy level of BESS $i$ |
| $\omega_i^0$ , $V_i^0$ | The nominal frequency and voltage of BESS $i$ |
| $u_i^{\omega}$ , $u_i^P$ , $u_i^E$ , $u_i^V$ and $u_i^Q$ | The consensus inputs of BESS $i$ |
| $\mathcal{G}$ , $\mathcal{V}$ , $\mathcal{E}$ , $\mathcal{W}$ | The communication graph and its vertex set, edge set, adjacency matrix with weights |
| $e_{ij}$ , $w_{ij}$ | The edge and weight between BESS $i$ and $j$ |
| $\mathcal{N}_i$ , $d_i$ . | The neighbor set of BESS $i$ and its cardinality |
| $\mathbf{0}$ , $A_0$ | The leader and its adjacency matrix |
| $\mathcal{E}_k$ , $\delta(\mathcal{E}_k)$ | The edge set in the $k^{th}$ iteration and its variance |
| $T_p$ , $T_{\omega}$ | The finite time of consensus |
| $\Delta P_i(t)$ , $\Delta\omega_i(t)$ | The real-time errors of the active power and the frequency |
| $W^{\tau}$ , $\tau_{ij}$ | The weight matrix with the delay time and the delay time between BESS $i$ and $j$ |
| BESS | Battery Energy Storage System |
| $i$ , $j$ | The index of agents |
| $v_{odi}$ , $v_{oqi}$ , $i_{odi}$ , $i_{oqi}$ | The d-axis and q-axis of voltage and current off BESS $i$ |
| $P_i$ , $Q_i$ | The calculated value of BESS $i$ |
| $\omega_{ci}$ | The cutoff frequency of the low-pass filter of BESS $i$ |
| $K_i^P$ , $K_i^Q$ | The droop coefficients of BESS $i$ |
| $\omega_i$ , $V_i$ , $P_i$ , $Q_i$ , $E_i$ | The frequency, voltage, active/reactive power and energy level of BESS $i$ |
| $\omega_i^0$ , $V_i^0$ | The nominal frequency and voltage of BESS $i$ |
| $u_i^{\omega}$ , $u_i^P$ , $u_i^E$ , $u_i^V$ and $u_i^Q$ | The consensus inputs of BESS $i$ |
| $\mathcal{G}$ , $\mathcal{V}$ , $\mathcal{E}$ , $\mathcal{W}$ | The communication graph and its vertex set, edge set, adjacency matrix with weights |
| $e_{ij}$ , $w_{ij}$ | The edge and weight between BESS $i$ and $j$ |
| $\mathcal{N}_i$ , $d_i$ . | The neighbor set of BESS $i$ and its cardinality |
| $\mathbf{0}$ , $A_0$ | The leader and its adjacency matrix |
| $\mathcal{E}_k$ , $\delta(\mathcal{E}_k)$ | The edge set in the $k^{th}$ iteration and its variance |

$T_p, T_\varpi$ The finite time of consensus

## I. Introduction

NOWADSYS, for the cooperative autonomous operation [1], multi-microgrid is widely applied in the modern electric power system, which contains generation sources, energy storage devices and loads [2]. What's more, the energy storage device has the capability of peak shaving and fault tolerance which contribute to the power-supply quality and reliability [3]. Meanwhile, the energy storage device incorporated into the multi-microgrid system can cope with the uncertainty brought by the generation of various kinds of renewable sources [4-8]. In all kinds of energy storage systems, battery energy storage system has been commonly implemented [9].

In the conventional centralized approach, it requires a central controller with a high bandwidth and communicates with all microgrids over a bidirectional and connected network [10-11]. Therefore, the central method is vulnerable to the single point-of-failure, which fails to deal with changeable and scalable communication structure [12]. Hence, the centralized way suffers from reliability and flexibility [13]. Alternatively, the multi microgrids working in the distributed fashion only need to make use of the information from themselves and their neighbors through a sparse communication topology and can cope with the load change [14-16]. In lots of papers, the distributed manner has been utilized to solve the energy management [17-21] and the distributed generation [22-23]. For example, [21] proposes a distributed method for the comprehensive utilization of multi energy sources in the electric power system, and [23] adopts the distributed gradient algorithm to seek for the online optimal generation.

Recently, the multi-agent system is adopted in distributed manner to coordinate the multi-microgrid system. Based on a distributed multi-agent cooperative control system, the power sharing among all energy storages reached balance and the frequency restored to the nominal value in [24]. For the balance of the State-of-Charge (SoC) in the multi BESSs, [25-26] proposed a distributed control strategy based on the multi-agent system, which could synchronize the SoC levels of all BESSs by using their neighbors' and themselves' information. Besides, [27] designed a secondary controller for the voltage, which transformed the secondary voltage control to a linear second-order tracker synchronization problem.

For a large-scale power system, the consensus among all BESSs is definitely affected by the location feature of each one. However, in all of above papers, the location information of each generation system has not been taken into consideration explicitly from the point of consensus coordination. And the heterogeneous nature of each battery has not introduced in [28-29], though the balance of SoC levels of multi BESSs in grid-connection mode and the restoration controller had been studied in [30] and [31] respectively. What's more, most of the existed studies are concentrating on the global asymptotic stability of the energy storage devices. That is, the states of BESS could not be stabilized in the finite time. In case of completeness, droop control is a commonly method to maintain the voltage and frequency in the micro-grid [32-34], will be kept again in this paper.

To address the above issues, a distributed step-by-step finite-time consensus control design is proposed, in which both the location and the heterogeneous nature are considered. Each state of all BESSs can be coordinated to supply power, and to facilitate the grid-connected, the voltage and frequency should be restored to the nominal value. So, for convenience in the switching between the grid-connected and island mode, we introduce the concept of "consensus" for the active/reactive power, the voltage, the frequency and the energy level. In the control structure, the hierarchical control method with droop characteristics is adopted. Based on the multi-agent system, each BESS communicates with each other over a sparse communication network. For convenience of grid connection, the voltage and the frequency of the multi-microgrid system should be regulated to the nominal values of the main grid. Therefore, a virtual leader is introduced into the control design. By applying the five inputs designed, the consensus of the active/reactive power sharing, the energy level, the voltage and the frequency of each BESS can be achieved within the finite time, and the voltage and the frequency can restore to the nominal values. The whole system can operate in both the grid-connected and island mode. The main contributions of this paper are presented as follow

1) For a large-scale multi-microgrid system, we consider the geographical location information of all BESSs, based on which BESSs are classified to several clusters. This has not been considered explicitly, especially in consensus coordination, in the previous research.
2) A distributed step-by-step finite-time control protocol with a virtual leader is proposed, by which all BESSs in each cluster can be coordinated. Therein, distance between each pair of BESSs is given special attention, and is normalized into the corresponding weight in the adjacency matrix. Meanwhile, all clusters can reach consensus, which could not be achieved in the previous research.
3) The approach proposed is completely distributed and only information from the neighbor needs to be collected. Compared with the centralized manner, the dependency on complex computation and high bandwidth is relaxed.
4) Communication delay tolerance and disturbance rejection are improved a lot in the suggested scheme, which are the benefits of the distributed consensus and the finite time coordination.

The rest sections of this paper are arranged as follow. At first, the heterogeneous battery energy storage system is introduced in Section II. Then, the preparatory knowledge and the input design of the distributed step-by-step finite-time consensus controller are presented in Section III. After that, the case studies designed for proposed control protocols are validated in Section IV. Finally, this paper is concluded in Section V.

## II. Heterogeneous Battery Energy Storage System (BESS)

Generally speaking, a BESS is consisted of an energy storage unit, an inverter, an output filter, a grid filter and the controller

[35]. Therein, the hierarchical control method is adopted in the control system, which includes a power calculator, a voltage controller, a current controller and a droop controller [36]. In addition, vector control is applied to regulate the voltage. In fact, the voltage controller and the current controller response faster enough than the droop controller, which can be neglected reasonably [37-38]. Therefore, a single BESS system can be simplified and expressed in Fig.1.

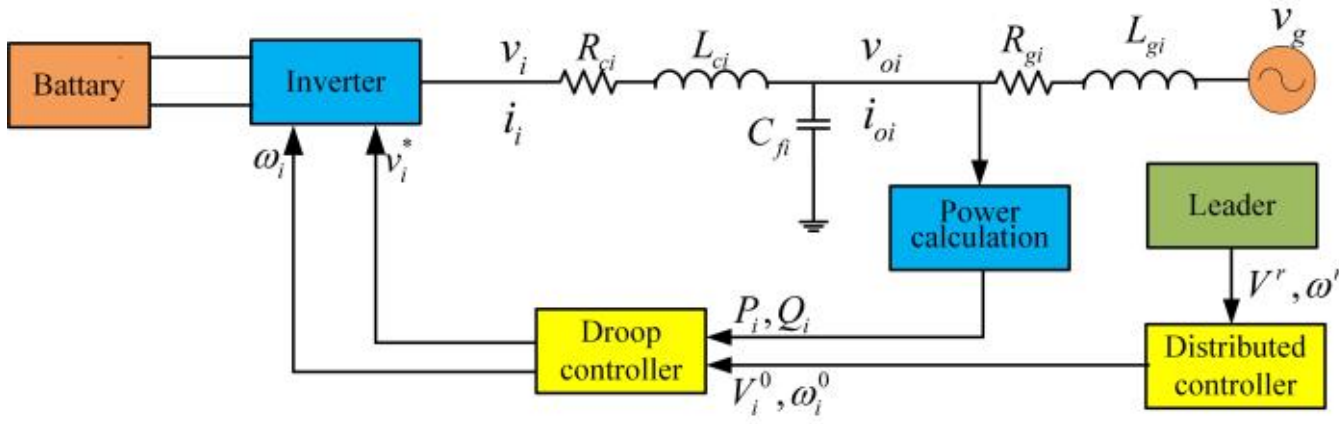

Fig. 1. The simplified structure of BESS $i$

By using the Park transformation, it is not difficult for us to get the active/reactive power, here we just listed them as Equations (1)-(2)

$$P_i = (v_{odi} i_{odi} + v_{oqi} i_{oqi}) \frac{\omega_{ci}}{s + \omega_{ci}} \tag{1}$$

$$Q_i = (v_{odi} i_{oqi} - v_{oqi} i_{odi}) \frac{\omega_{ci}}{s + \omega_{ci}} \tag{2}$$

By locally using the measured values of the active/reactive power, the droop controller is in charge of the stability of frequency and voltage, can be expressed by (3) [35],

$$\begin{bmatrix} \omega_i \\ V_i \end{bmatrix} = \begin{bmatrix} 1 & -1 & 0 & 0 \\ 0 & 0 & 1 & -1 \end{bmatrix} \begin{bmatrix} \omega_i^0 \\ K_i^P P_i \\ V_i^0 \\ K_i^Q Q_i \end{bmatrix} = \begin{bmatrix} 1 & -1 & 0 & 0 \\ 0 & 0 & 1 & -1 \end{bmatrix} \begin{bmatrix} \omega_i^0 \\ P_i \\ V_i^0 \\ Q_i \end{bmatrix} \tag{3}$$

Where $K_i^P P_i$ and $K_i^Q Q_i$ are denoted by $P_i$ and $Q_i$ correspondingly. In addition, voltage amplitude can be obtained as (4) under the $dq$ frame.

$$V_i = \sqrt{v_{odi}^2 + v_{oqi}^2} \tag{4}$$

For brevity, quadrature components of $v_{oi}$ is set as zeros, i.e. $v_{oqi}^* = 0$, and voltage amplitude $V_i$ can be replaced with $v_{odi}^*$, i.e. $V_i = v_{odi}^*$. In order to synchronize all BESS, five inputs are designed and the control system can be presented in (5)-(6), i.e. $u_i^{\omega}$, $u_i^P$, $u_i^E$, $u_i^V$ and $u_i^Q$ [35]. Inspired by [39], this paper introduces the sate-of-charge to measure the energy level.

Due to the usage of the character different droop coefficients, the energy level regulation has heterogeneous nature in this sense. And the distributed control scheme of the single BESS can be shown as Figure 2.

$$\begin{bmatrix} \dot{\omega}_i^0 \\ \dot{P}_i \\ \dot{E}_i \\ \dot{V}_i^0 \\ \dot{Q}_i \end{bmatrix} = \begin{bmatrix} 0 & 0 & 0 & 0 & 0 \\ 0 & 0 & 0 & 0 & 0 \\ 0 & \frac{-1}{3600} & 0 & 0 & 0 \\ 0 & 0 & 0 & 0 & 0 \\ 0 & 0 & 0 & 0 & 0 \end{bmatrix} \begin{bmatrix} \omega_i^0 \\ P_i \\ E_i \\ V_i^0 \\ Q_i \end{bmatrix} + \begin{bmatrix} u_i^{\omega} \\ u_i^P \\ u_i^E \\ u_i^V \\ u_i^Q \end{bmatrix} \tag{5}$$

$$\begin{bmatrix} \omega_i \\ P_i \\ E_i \\ V_i \\ Q_i \end{bmatrix} = \begin{bmatrix} 1 & -1 & 0 & 0 & 0 \\ 0 & 1 & 0 & 0 & 0 \\ 0 & 0 & 1 & 0 & 0 \\ 0 & 0 & 0 & 1 & -1 \\ 0 & 0 & 0 & 0 & 1 \end{bmatrix} \begin{bmatrix} \omega_i^0 \\ P_i \\ E_i \\ V_i^0 \\ Q_i \end{bmatrix} \tag{6}$$

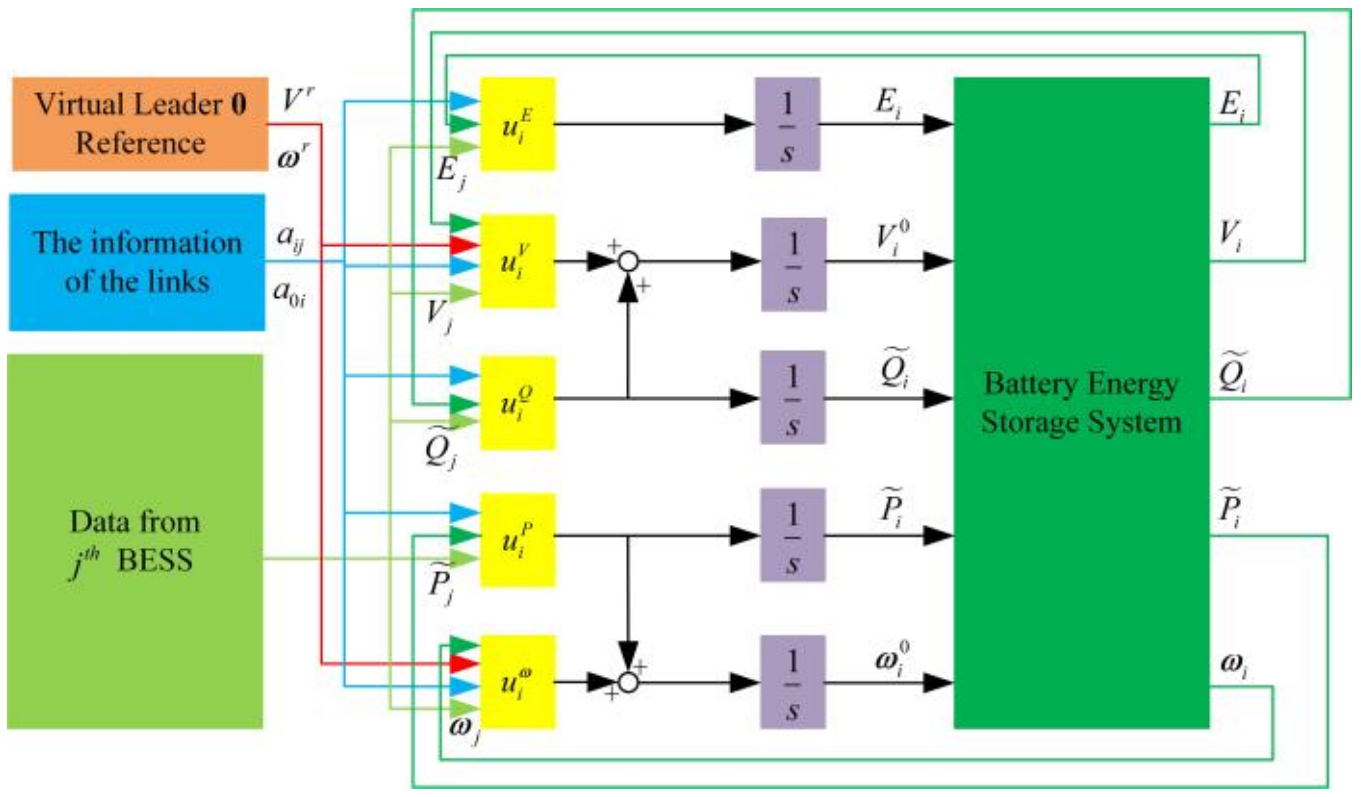

Fig. 2. The distributed consensus control structure of BESS $i$

## III. Distributed Approach for Energy Management Based on Multi-Agent

Based on multi-agent system, this paper designs a step-by-step fine-time consensus algorithm to synchronize battery energy level, active/reactive power sharing, voltage magnitude and frequency of each BESS. And a virtual leader is introduced to regulate the voltage magnitude and frequency to the reference values.

### A. Preliminaries

In a communication graph with weights $\mathcal{G}(\mathcal{V}, \mathcal{E}, \mathcal{W})$ of the multi-agent system, $\mathcal{V} = \{v_i \mid i \in [1, 2, \cdots, n]\}$ and $\mathcal{E} = \{e_{ij} = (i, j) \mid v_i \in \mathcal{V}, v_j \in \mathcal{V}\}$ are the set of vertexes and edges respectively, and $\mathcal{W} = (w_{ij})_{n \times n}$ is the weight matrix. If agent $i$ is connected to agent $j$, $w_{ij} > 0$ and agent $j$ is the neighbor of agent $i$; otherwise, $w_{ij} = 0$. Therefore, $\mathcal{W}$ is considered as the adjacency matrix associated with $\mathcal{G}$ [40]. The neighbor set of agent $i$ is denoted by $\mathcal{N}_i$ with the cardinality $d_i$.

In the leader-follower algorithm, there is a virtual leader **0** in the communication structure, which is usually an external instruction. In order to save the hardware facilities, a small part of agents is connected to the virtual leader **0**, whose neighbor set is donated by $N_0$ with the cardinality $d_0$. In addition, $A_0=(a_{0i})_n$ represents the adjacency matrix of the virtual leader **0**. If the virtual leader **0** connects with agent $i$, $a_{0i}=1$; conversely, $a_{0i}=0$.

Kruskal is a kind of the minimum spanning tree algorithm, which runs by traversing the edge for each iteration. In this paper, we adopt a Clustering Algorithm based on Kruskal (CAK) to classify the distributed BESSs [41]. To seek for the maximum value of $\delta_k$ in Equation (7), a certain of edges in $E_k$ will be discarded and other edges make up the new set $E_{k-1}$ in the $k^{th}$ iteration. When the Inequality (8) is satisfied, iteration ends.

$$\delta_k=\left|\delta(E_k)-\delta(E_{k-1})\right| \tag{7}$$

$$\left|\delta(\varepsilon_k)-\delta(\varepsilon_{k-1})\right|<\left|\delta(\varepsilon_1)-\delta(\varepsilon_0)\right| \tag{8}$$

Besides, in the multi-micro-grid system, an isolated BESS will cause instability and it will be classified into the nearest cluster.

Considering the geographical location features, all agents can be divided into several clusters. Figure 3 is the diagram of the step-by-step consensus scheme. Therein, all agents are classified to two clusters F1 and F2. Moreover, L1 in Cluster 1 and L2 in Cluster 2 are considered as the active leaders which can receive the information from the external and other clusters, and L is a virtual leader from the outside. Accordingly, agents in F1 and F2 will follow L1 and L2 respectively.

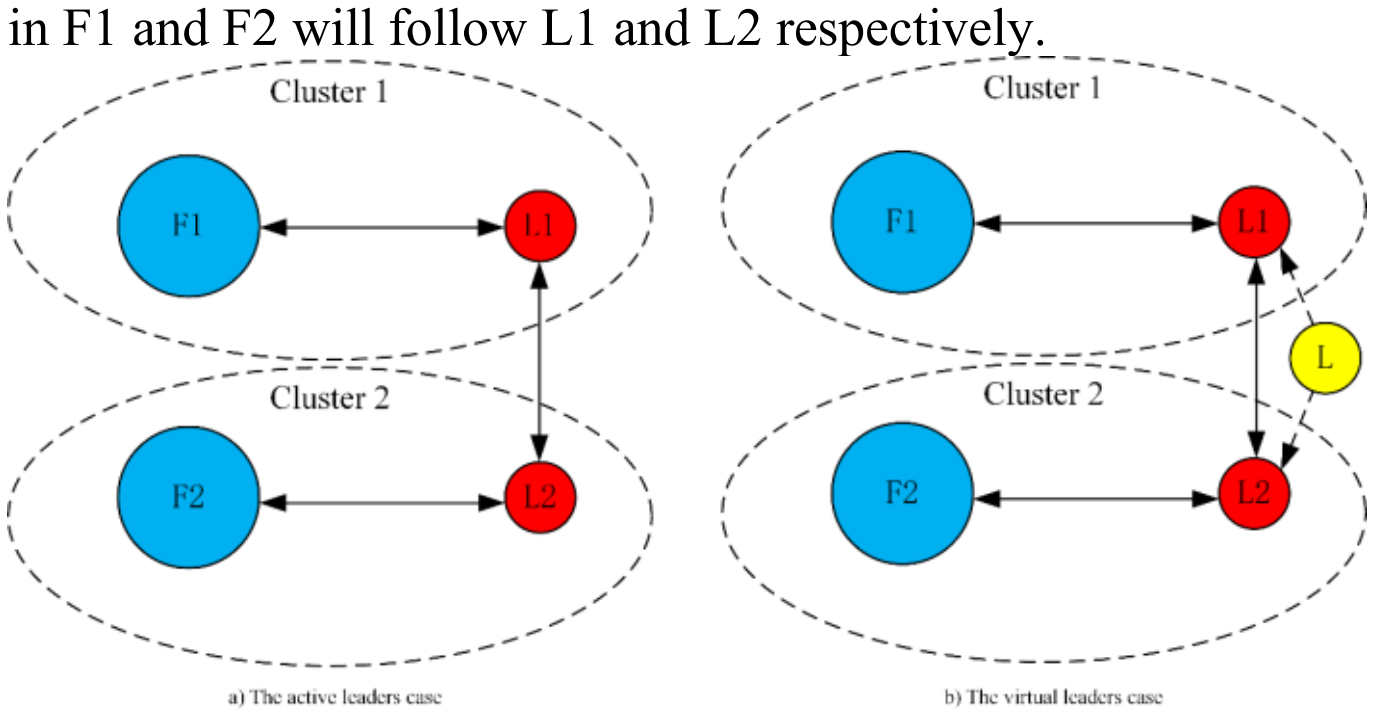

Fig. 3. The diagram of the step-by-step consensus algorithm, a) is the active leader case, and b) is the virtual leader case.

*B. The finite-time consensus input designed*

Inspired by [42], this section will be devoted to present the finite-time input designed, the proof will be provided and the formula for computation of the convergence time will also be made available.

**Theorem 1:** Active/reactive power and energy level will reach the average consensus when $t\ge T_p$, while the frequency and voltage will follow the states of the leader when $t\ge T_\omega$, respectively under the control laws from (9) to (13).

$$u_i^P=-\sum_{j=1}^{n}w_{ij}\,\mathrm{sgn}(P_i-P_j)\left|P_i-P_j\right|^\eta \tag{9}$$

$$u_i^Q=-\sum_{j=1}^{n}w_{ij}\,\mathrm{sgn}(Q_i-Q_j)\left|Q_i-Q_j\right|^\eta \tag{10}$$

$$u_i^E=-\sum_{j=1}^{n}w_{ij}\,\mathrm{sgn}(E_i-E_j)\left|E_i-E_j\right|^\eta \tag{11}$$

$$\begin{aligned}u_i^{\omega}=&-\sum_{i=1}^{n}w_{ij}\,\mathrm{sgn}(\omega_i-\omega_j)\left|\omega_i-\omega_j\right|^\eta\\&-\sum_{i=1}^{n}w_{ij}\,\mathrm{sgn}(P_i-P_j)\left|P_i-P_j\right|^\eta\\&-a_{0i}\,\mathrm{sgn}(\omega_i-\omega^r)\left|\omega_i-\omega^r\right|^\eta\end{aligned} \tag{12}$$

$$\begin{aligned}u_i^{V}=&-\sum_{i=1}^{n}w_{ij}\,\mathrm{sgn}(V_i-V_j)\left|V_i-V_j\right|^\eta\\&-\sum_{i=1}^{n}w_{ij}\,\mathrm{sgn}(Q_i-Q_j)\left|Q_i-Q_j\right|^\eta\\&-a_{0i}\,\mathrm{sgn}(V_i-V^r)\left|V_i-V^r\right|^\eta\end{aligned} \tag{13}$$

Where $\mathrm{sgn}(\bullet)$ is the symbolic function of the concerned, $\eta$ is an adjustable parameter and $0<\eta<1$, $\omega^r$ and $V^r$ are the reference value of frequency and voltage which are included in the virtual leader. Here, we provide the proof of the finite-time consensus and the calculation of consensus time. In order to give the proof of **Theorem 1**, we need to provide some necessary preparations at first. **Lemma 1** to **Lemma 4** are listed as follows to serve for the preparation of Theorem 1 proof.

**Lemma 1** [43]: If $y_1,y_2,\cdots,y_n\ge 0$ and $0<r<p$, then

$$\left(\sum_{i=1}^{n}y_i^p\right)^{\frac{1}{p}}\le\left(\sum_{i=1}^{n}y_i^r\right)^{\frac{1}{r}} \tag{14}$$

**Lemma 2** [44]: Assume that $L$ is a Laplacian matrix, and $\lambda_b$ is the smallest positive eigenvalue of $L$, then we have

$$x^TLx=\frac{1}{2}\sum_{i,j=1}^{n}a_{ij}(x_j-x_i)^2 \tag{15}$$

$$x^TLx\ge\lambda_b x^Tx \tag{16}$$

Where $a_{ij}$ is the element of the adjacency matrix $A$ of $L$

**Lemma 3** [45]: Assume that $V(x)$ is a C-regular function for $R^n\to R$, and $x(t)$ is a continuous function of $t$ for $[0,+\infty)\to R^n$. If the inequality (17) is satisfied, then $V(x)$ will converge to 0 when $t\ge T$.

$$\frac{dV(t)}{dt}\le -KV^{\alpha}(t) \tag{17}$$

Where $K>0$, and $0<\alpha<1$. And $T$ is estimated by (18)

$$T=\frac{V^{1-\alpha}(0)}{K(1-\alpha)} \tag{18}$$

**Lemma 4** [11]: Assume that $A_0$ is a semi-positive definite diagonal matrix, and $L_1$ is a Laplacian matrix, then $(A_0+L_1)$ is also a positive semidefinite matrix, and

$$x^T(A_0+L_1)x=\sum_{i=1}^{n}a_{0ij}x_i^2+\frac{1}{2}\sum_{i,j=1}^{n}a_{1ij}(x_i-x_j)^2 \tag{19}$$

Suppose that $\lambda_c$ is the smallest positive eigenvalue of $(A_0+L_1)$, then

$$x^T(A_0+L_1)x\ge\lambda_c x^Tx \tag{20}$$

With **Lemmas 1** to **4** in hand, we can start the proof of **Theorem 1.** And in case of the convenience of reading, the proof of Theorem 1 is shifted to APPENDIX.

Finally, when $0<\eta<1$, the reactive power and the energy level also can reach average consensus, and the voltage amplitude will track the leader.

Actually, for $\eta=0$, [46] shows that the voltage and frequency still can converge to the leader, but the real-time error becomes discontinuous. For $\eta=1$, that is the gradual consistency case studied in [35]-[36]. What' more, the finite consensus can be ensured by utilizing the distributed information on the condition of the fixed communication structure. Even if the communication structure is changed, or small amount of encounters communication failure, all BESSs will still be synchronized correspondingly. However, the convergence time varies under the different communication graph. So, only if the communication topology switch to the sparse one in time, this algorithm is robust to the link changes.

### *C. The step-by-step finite-time consensus controller considering the delay time*

In the communication graph, there are usually some delays. In this section, we will design the controller considering the communication delay time $\tau_{ij}$ between agent $i$ and agent $j$. According to [47], the weight matrix based on the delay time is improved as (34)

$$W^{\tau}(i,j)=\begin{cases}0, & j=i\\ e^{-\tau_{ij}s}w_{ij}, & j\in N_i\\ 0, & other\end{cases} \tag{34}$$

## IV. Simulation Result

In this paper, the step-by-step finite-time consensus controller proposed only exploits the local information in distributed manner, which contributes to coordinate all heterogeneous BESSs. As a result, whatever each BESS is at high or low level initially, they can be consumed or charged simultaneously. At the same time, frequency and voltage can be regulated to 1 *p.u.* which is the necessary condition for the grid-connection mode, and the total output power can meet the load demand. So, in this section, 4 delicate cases will be employed to test the effect of the controller. The modified IEEE 57-bus system shown in Fig. 4 is adopted for the case studies, which contains 7 generators and 42 load buses. The load distribution is obtained from [48]. And over undirected communication networks as shown in Fig. 5, the step-by-step finite-time consensus protocols are applied among all BESSs. Therein, we consider the distance between each pair of BESSs and normalize them as the weights in the adjacency matrix, then divide them into several categories as shown in the Fig. 5 a3) and b3) by CAK. The specific values of droop gains are presented in Table I. All of case studies are implemented by the Matdyn [49], which is a MATLAB toolbox based on MATPOWER [50] and obtained from [51].

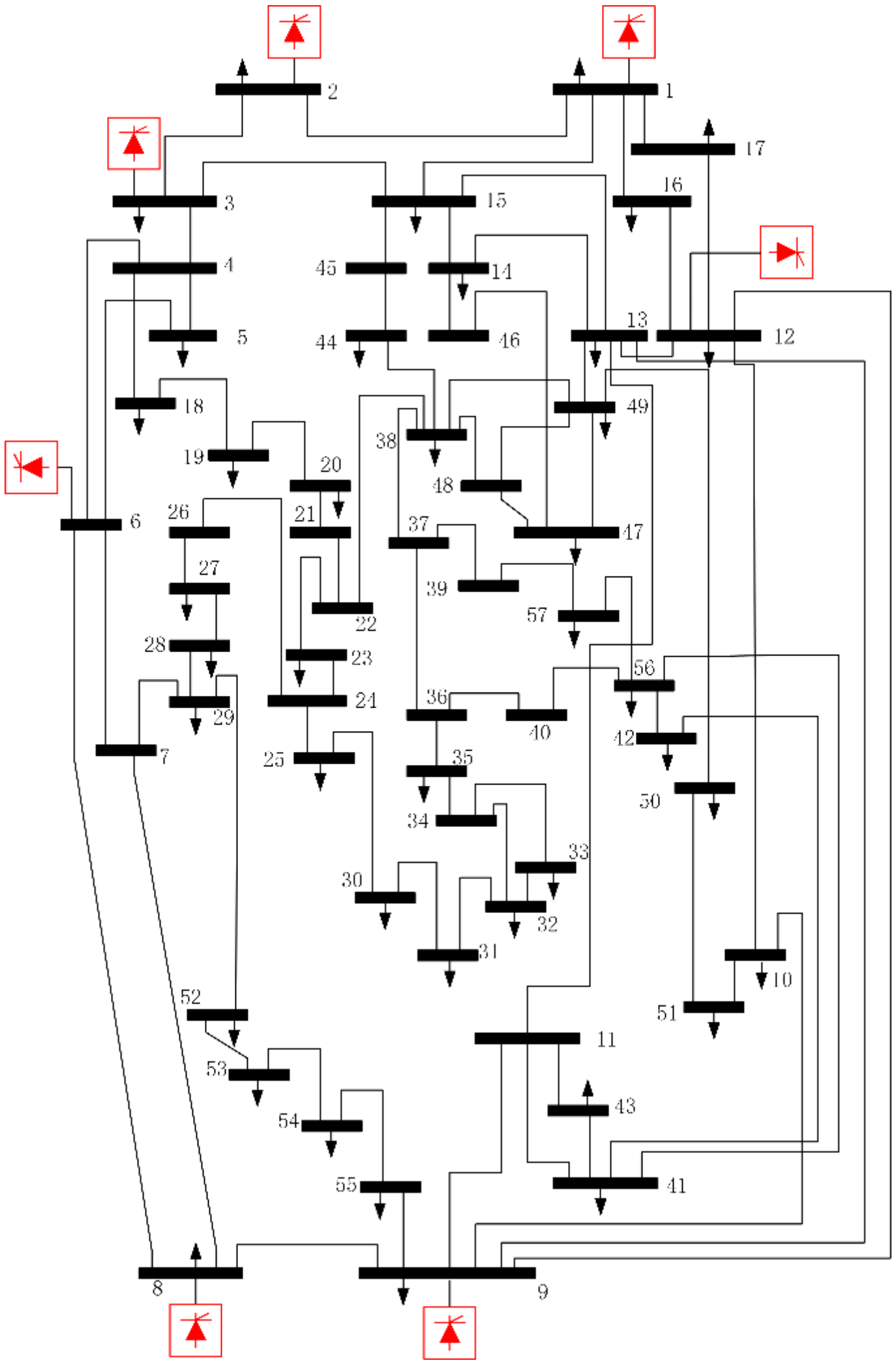

Fig. 4. The modified IEEE 57-bus system

TABLE I
The parameters of the test system adopted

| Par. | Value | Par. | Value |
|---|---|---|---|
| $K_1^P$ | 1 | $K_1^Q$ | 1 |
| $K_2^P$ | 1.2 | $K_2^Q$ | 1.21 |
| $K_3^P$ | 1.25 | $K_3^Q$ | 1.3 |
| $K_4^P$ | 0.9 | $K_4^Q$ | 1.23 |
| $K_5^P$ | 0.95 | $K_5^Q$ | 0.95 |
| $K_6^P$ | 1.02 | $K_6^Q$ | 1.06 |
| $K_7^P$ | 1.1 | $K_7^Q$ | 1.13 |
| $\omega^r$ | 1 *p.u* | $V^r$ | 1 *p.u* |
| $\eta$ | 0.5 | | |

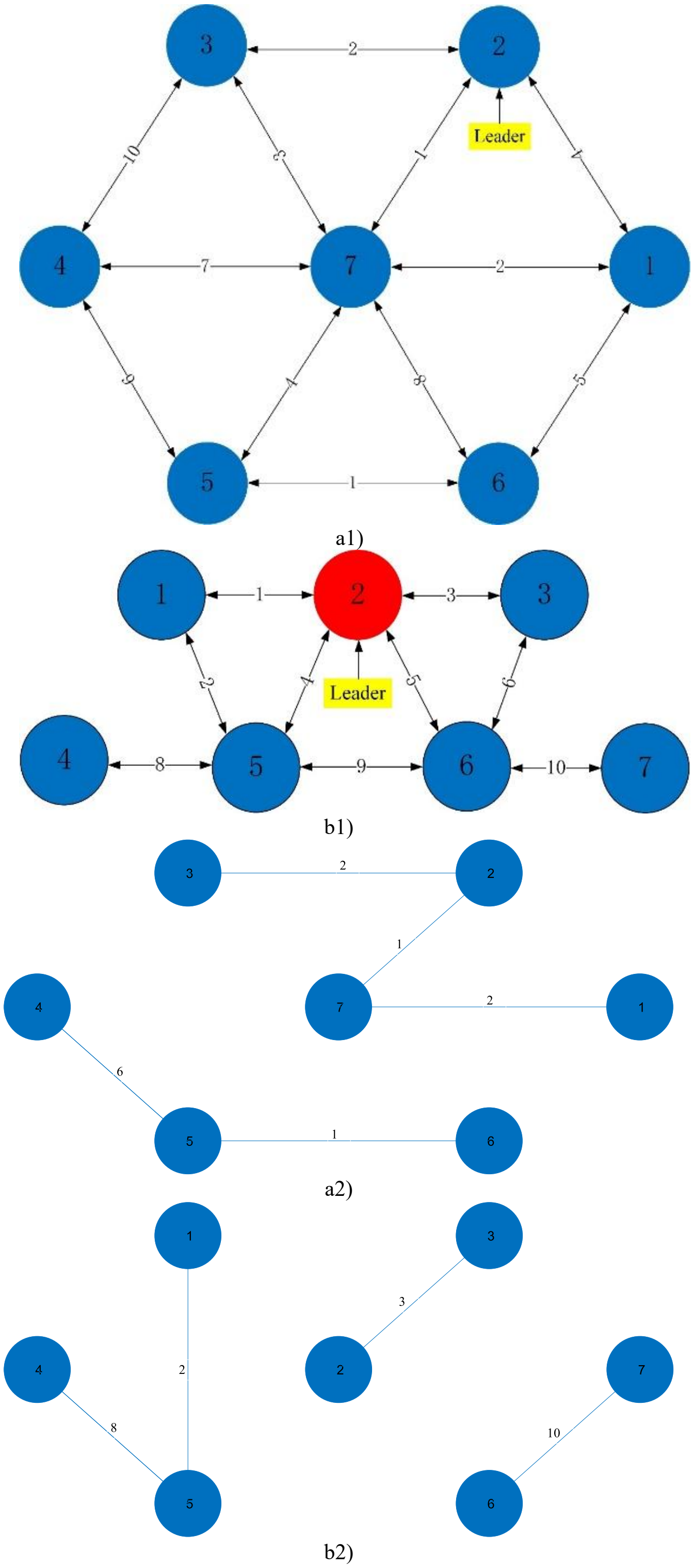

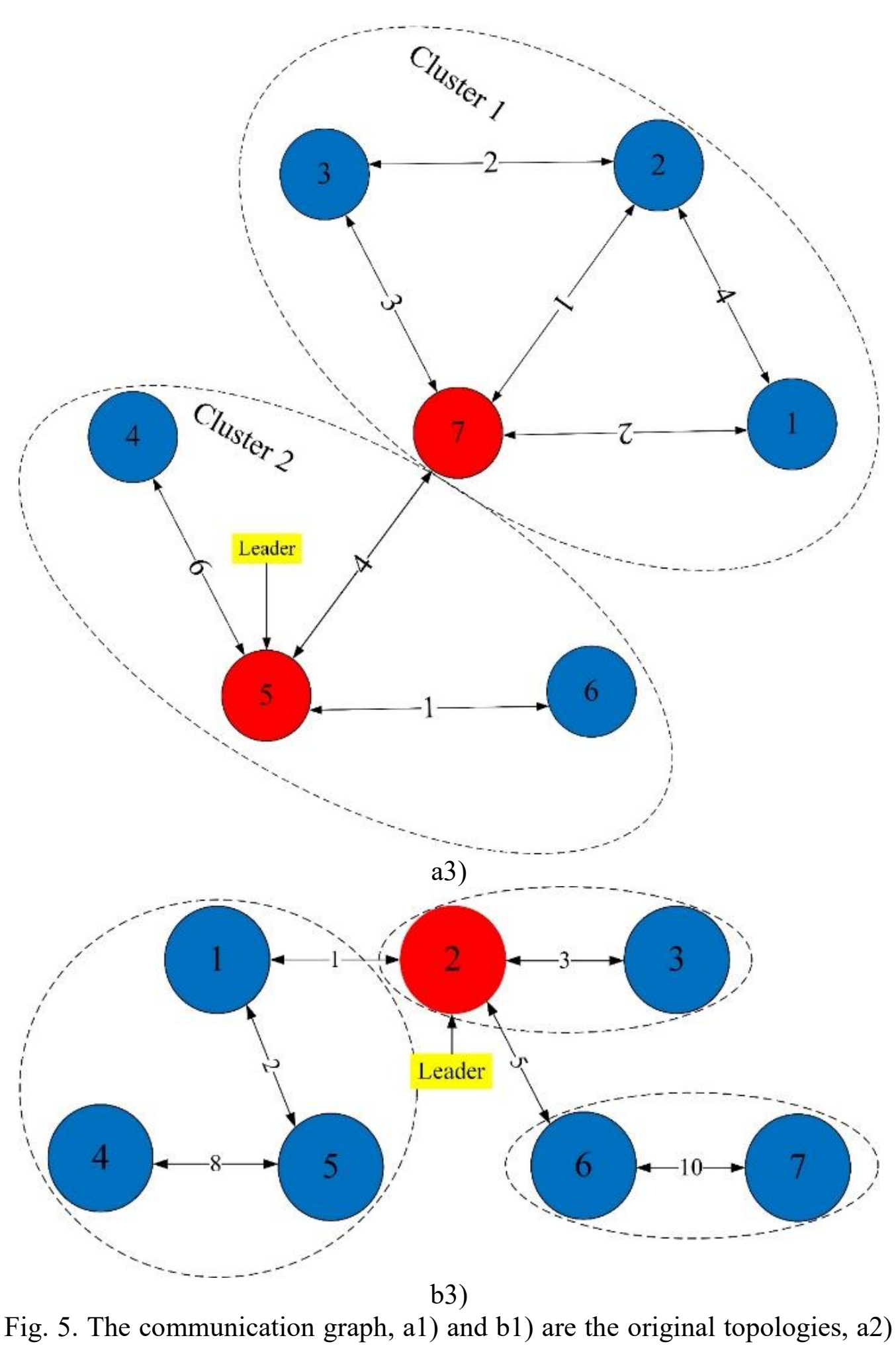

Fig. 5. The communication graph, a1) and b1) are the original topologies, a2) and b2) are the clustering results of a1) and b1) based on CAK, a3) and b3) are the topologies considering the location information.

### A. *Case 1: Performance of the Proposed Approach*

This case is designed to test the performance of the proposed consensus controller. The test system operates in the island mode, and all BESSs are working at discharge fashion. Multi BESSs communicate with each other over the topology shown in a3 of Figure 4. Under the action of the droop controller, all states of BESS start at steady at $t=0\text{s}$, and the controller is activated at $t=20\text{s}$. In addition, there is a $0.5+0.5j$ load decrease while a $1+1j$ load increase at $t=80s$. The simulation results are shown in Fig. 6 and Fig. 7. Therein, Fig. 6 shows the change rules of all states of 7 BESSs, and take the frequency and active power as two examples to display the finite convergence of the error system in Fig. 7.

In fact, the droop controller has the characteristics of discrepancy regulating for speed/frequency, so both voltage amplitude and frequency of each BESS deviate from their rated value values before $t=20s$. However, after the consensus controller is activated, the error systems of all BESSs converge to 0. That is, the active/reactive power of each BESS can reach average consensus to meet the supply-demand balance, and voltage and frequency can be restored to the nominal values. And energy level of each BESS can be coordinated by an additional input such that energy level balance and active power sharing can be reached independently.

Load switching is also studied in this case. It can be seen that, irrespective of the load changes, the active/reactive power sharing and the balance of energy level among all BESSs are ensured easily, and the frequency and voltage can be regulated to the nominal values quickly. So, the step-by-step finite-time consensus controller has the capability of plug-and-play and it is robust for load change in terms of the simulation results.

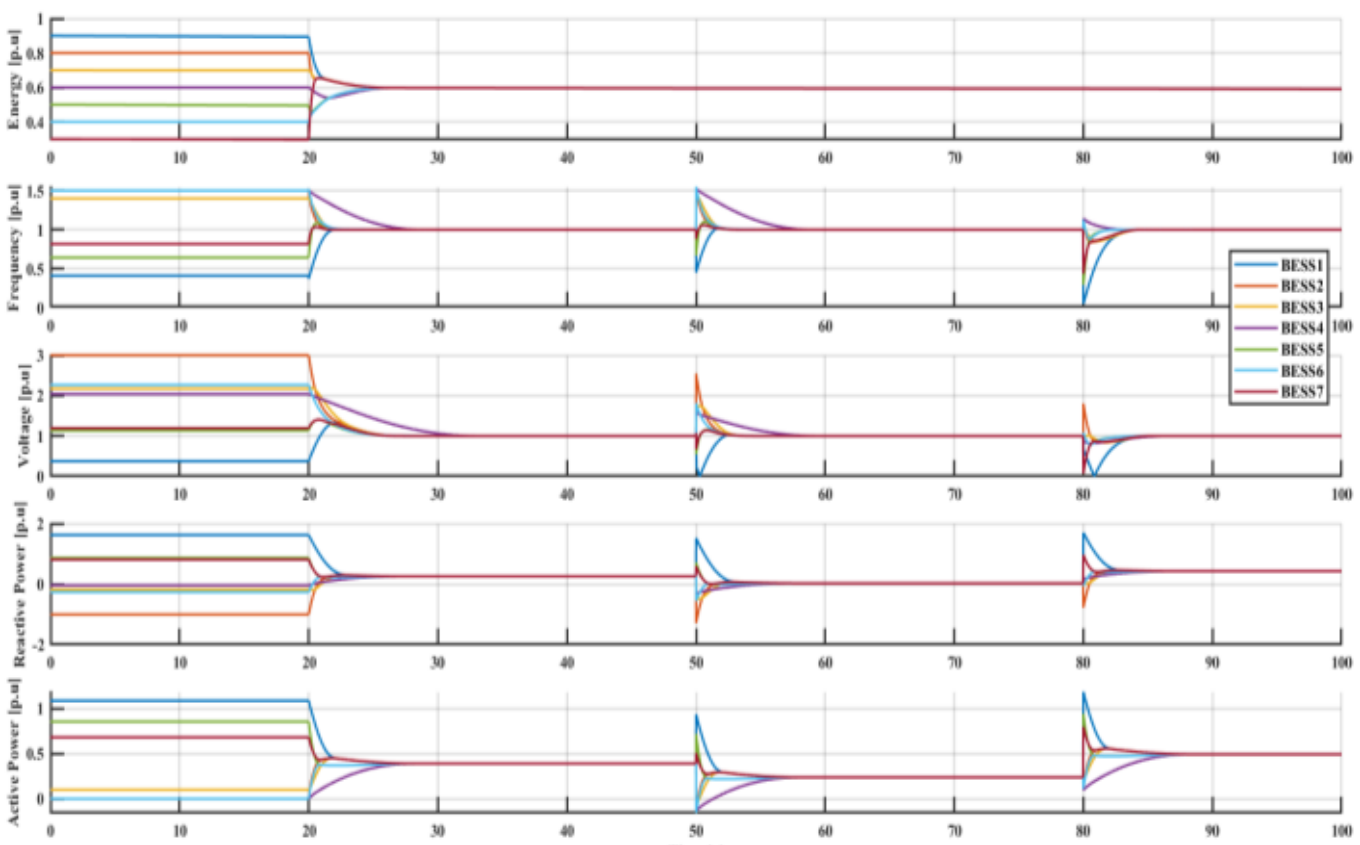

Fig. 6. The simulation results of all state variables of multi BESSs in Case 1

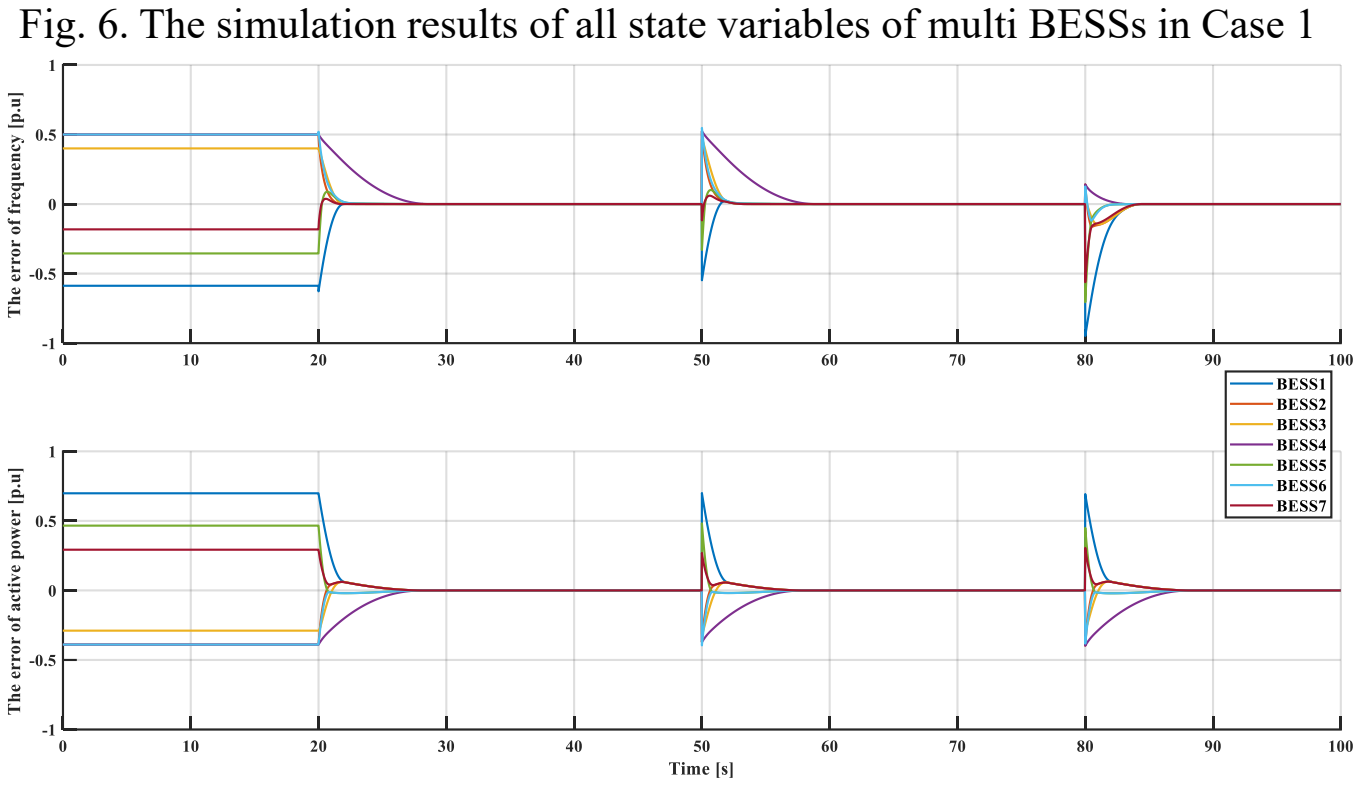

Fig. 7. The simulation results of the real-time errors of the frequency and active power sharing of each BESS in Case 1

## B. *Case 2: The influence of communication graph change*

In this case study, the performance in coping with the uncertain communication structure is validated while the multi BESSs are working in the islanded mode, i.e. the discharging mode at beginning. After the controller activated at $t = 20s$, the whole system is operating in the grid-connection mode, i.e. the charging mode, and the communication topology is under the a3 of Fig. 4. When the simulation system runs to $t = 50s$, the connectivity style switches from a3 to b3 of Fig. 4. Then a $0.5 + 0.5j$ load decrease takes place at $t = 80s$. Eventually, the real-time states of all BESSs are present in Fig. 8, and Fig. 9 is for the real-time error of the active power and frequency.

Obviously, the step-by-step finite-time consensus controller can operate in both islanded and grid-connected mode. And the diversity and time-varying characteristics of the link connection have no effect on the stability of the system. In addition, when a small amount of BESSs are isolated, only if remains are still connected, they can also be cooperated such that the whole system works reliably and stably.

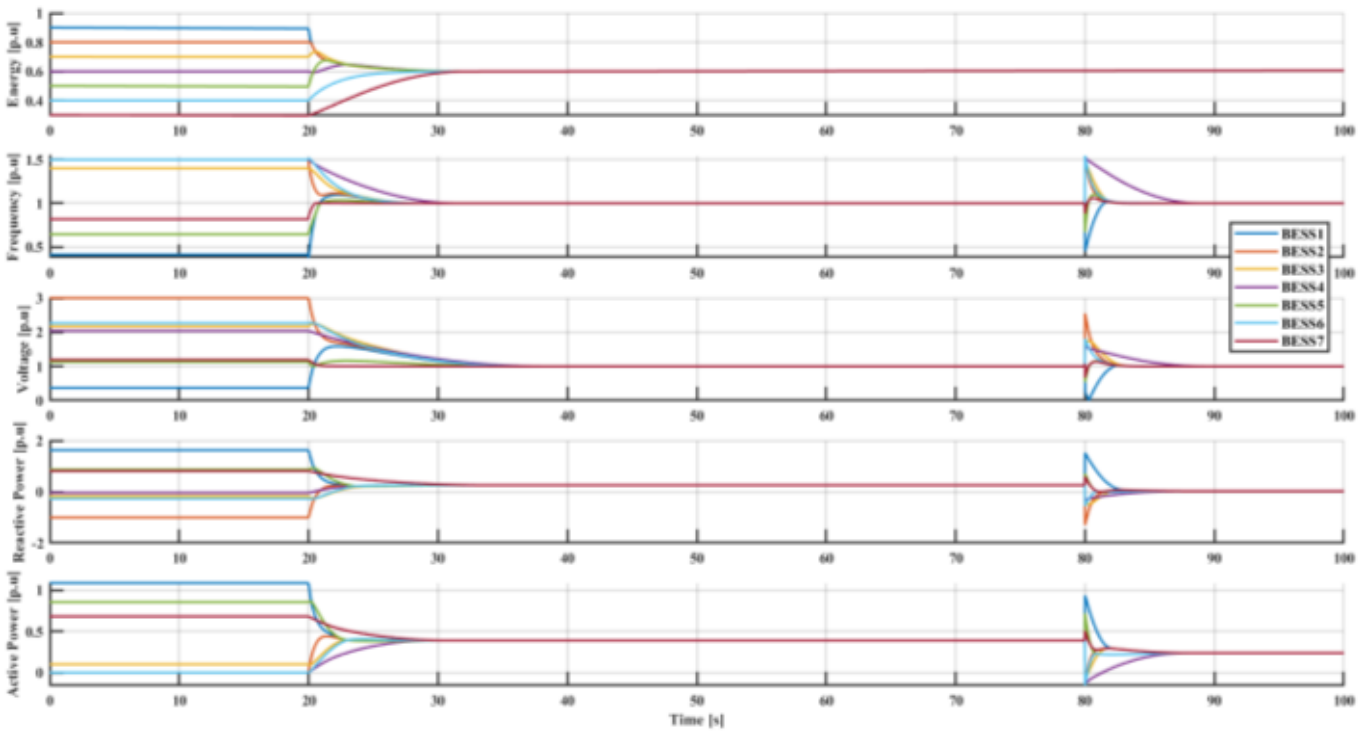

Fig. 8. The simulation results of all state variables of multi BESSs in Case 2

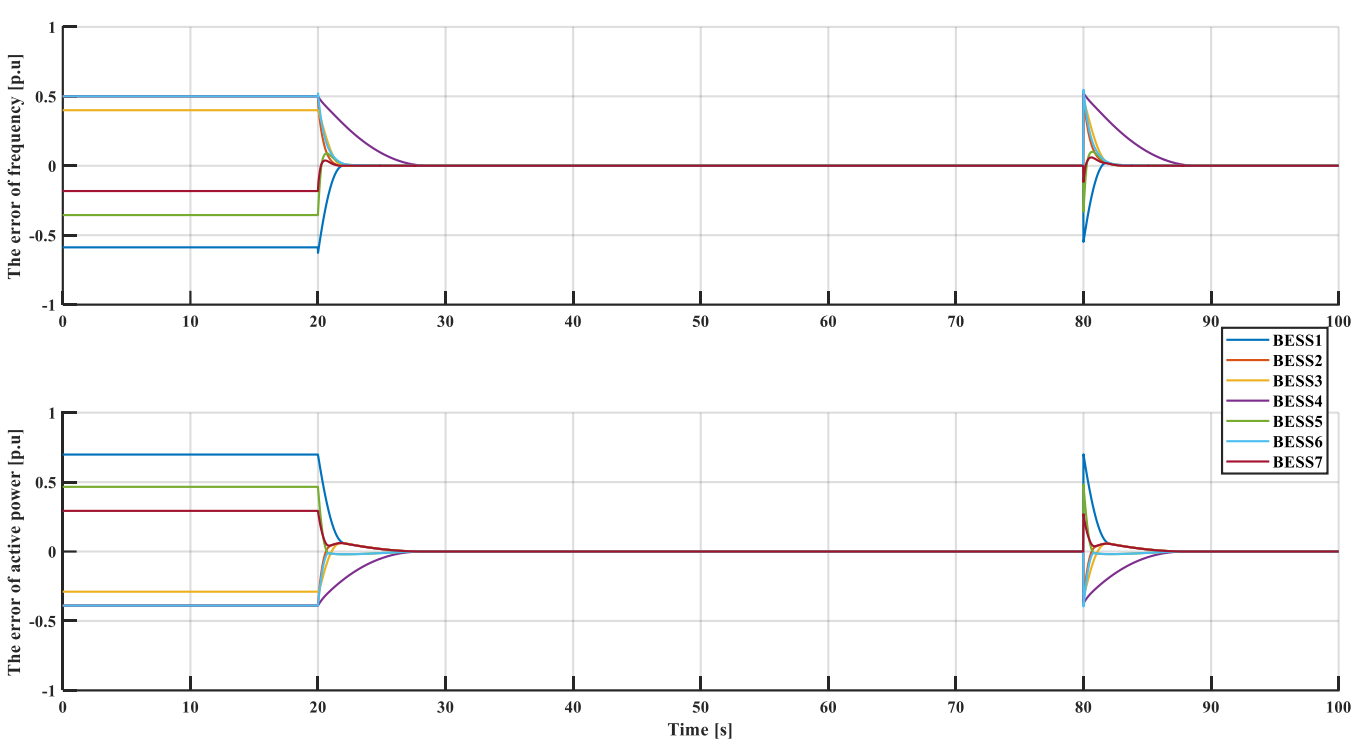

Fig. 9. The simulation results of the real-time errors of the frequency and active power sharing of each BESS in Case 2

## C. *Case 3: The effect of the adjustable parameter* $\eta$

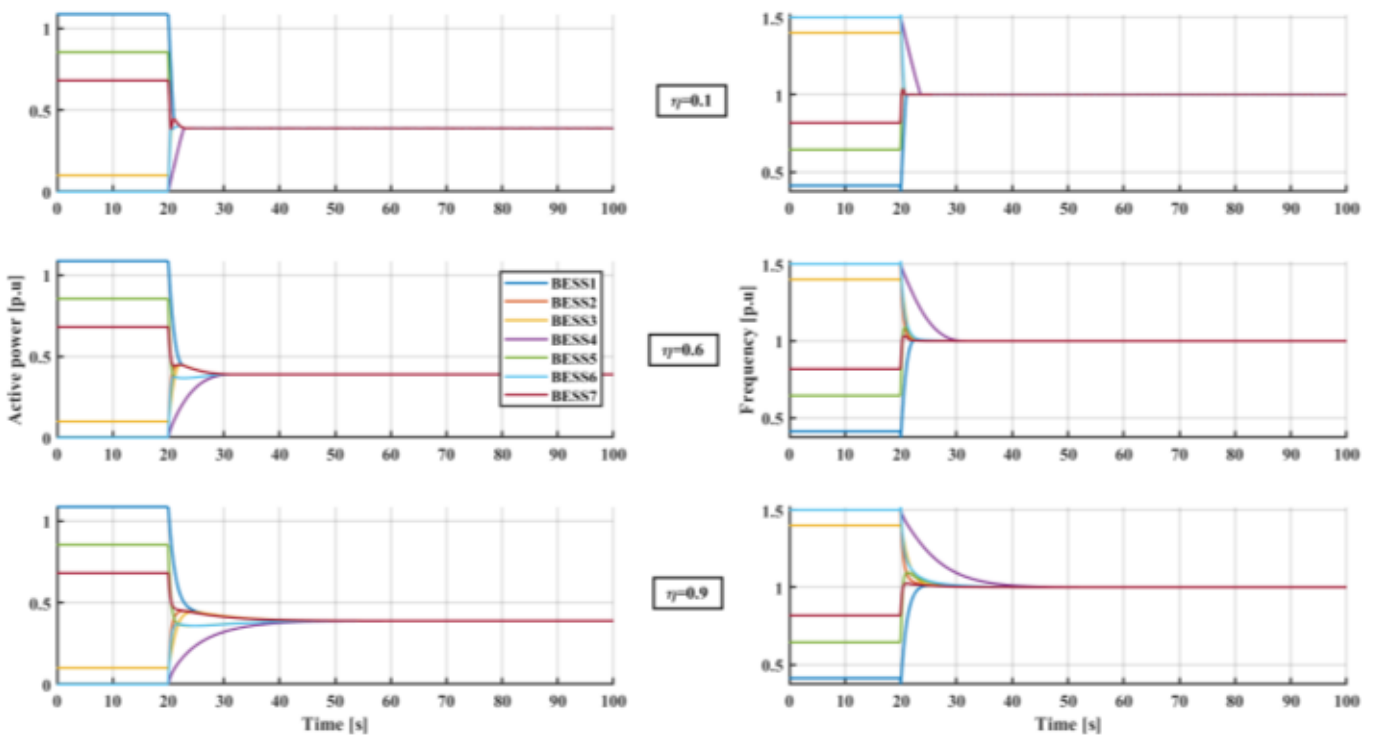

Fig. 10. The simulation results of all state variables of multi BESSs under different $\eta$ in Case 3

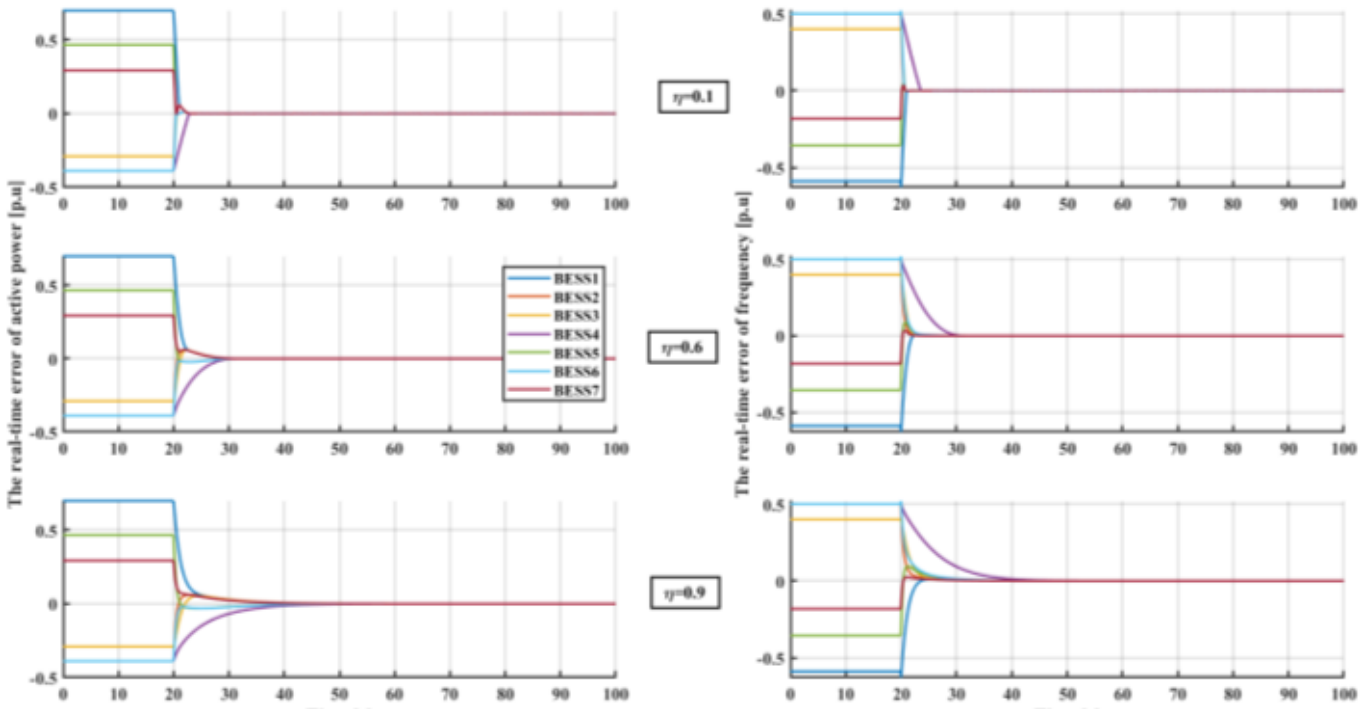

Fig. 11. The simulation results of the real-time errors of the frequency and active power sharing of each BESS under different $\eta$ in Case 3

In the previous section, the adjustable parameter $\eta$ is key to calculate the convergence time. Therefore, this case is arranged to test the effect on the finite time by set $\eta = -0.1, -0.6 \text{ and } -0.9$. Specially, different values of $\eta$ are considered in active power, frequency and corresponding error system. And the communication graph in this case is selected as a3 of Fig. 4. The simulation results are exhibited from Fig. 10 to Fig. 11.

It is interesting to check the relationship of speed convergence with the selection value of $\eta$. Under the condition of $-1 < \eta < 0$, the smaller the value of $\eta$, the slower the convergence of the active power and frequency of each BESS. Correspondingly, with the decrease of $\eta$, the error systems converge to zeros faster, while the performance of the system becomes more aggressive. So, we must weigh the system performance and convergence speed to choose a proper $\eta$.

### D. Case 4: The performance of the controller designed in dealing with the communication delay

This case is designed to cope with the communication delay. And the delay time $\tau_{ij}$ is set as 500ms, which exceeds the upper bound of the delay time required by UTC and Avista [51]. The system is operated in the island mode, and all BESSs communicate with each other over the topology of a3 of Fig. 4. When $t = 20s$, the consensus controller begins to work. In addition, there is a load $0.5 + 0.5j$ decrease at $t = 50s$. Finally, the simulation results are displayed in Fig.12-Fig.13.

It can be seen that although there are some oscillations, all states of BESSs still can reach consensus, even under the worsen cases of load changes. However, it should be pointed out that compared with Case 1, the time of the energy level and active power consensus among all clusters was delayed under the communication delay. Likewise, for the error system, the oscillations will occur in each cluster and it takes more time to converge to 0 for each cluster.

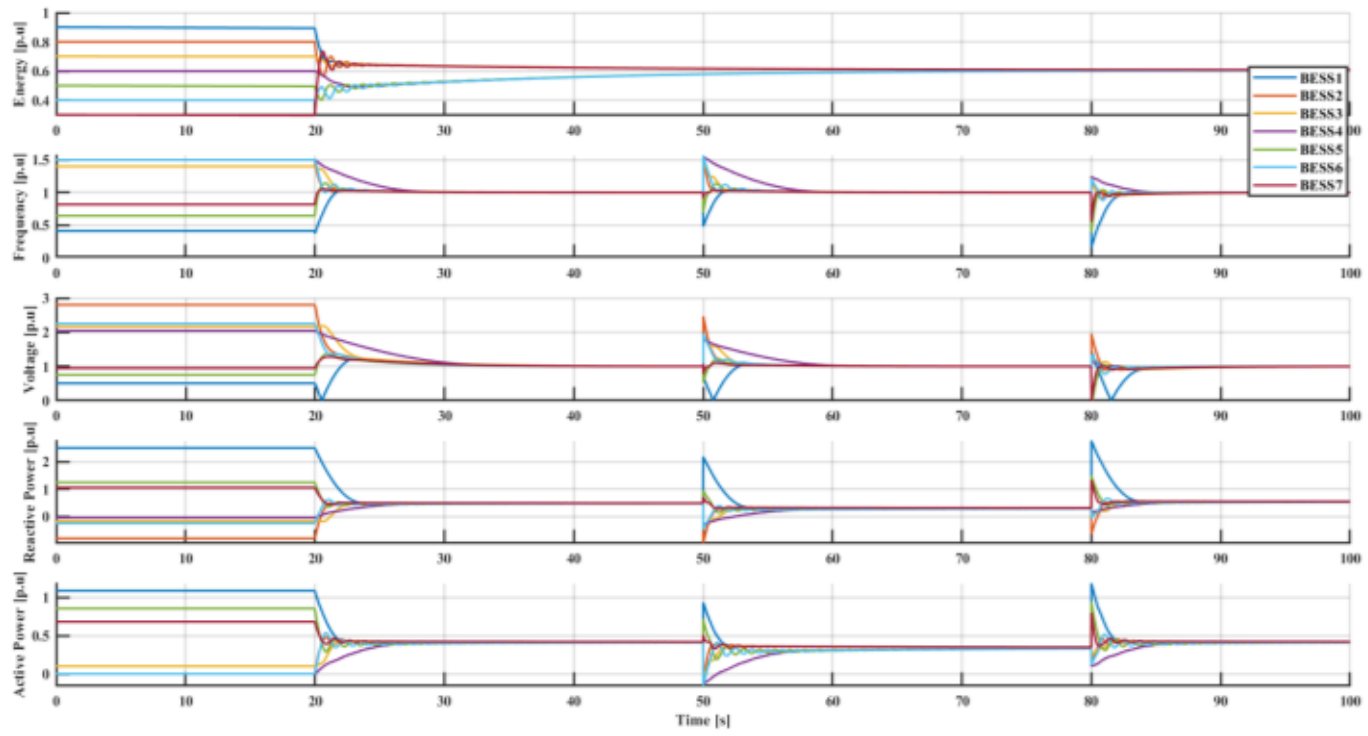

Fig. 12. The simulation results of all state variables of multi BESSs under the communication delay in Case 4

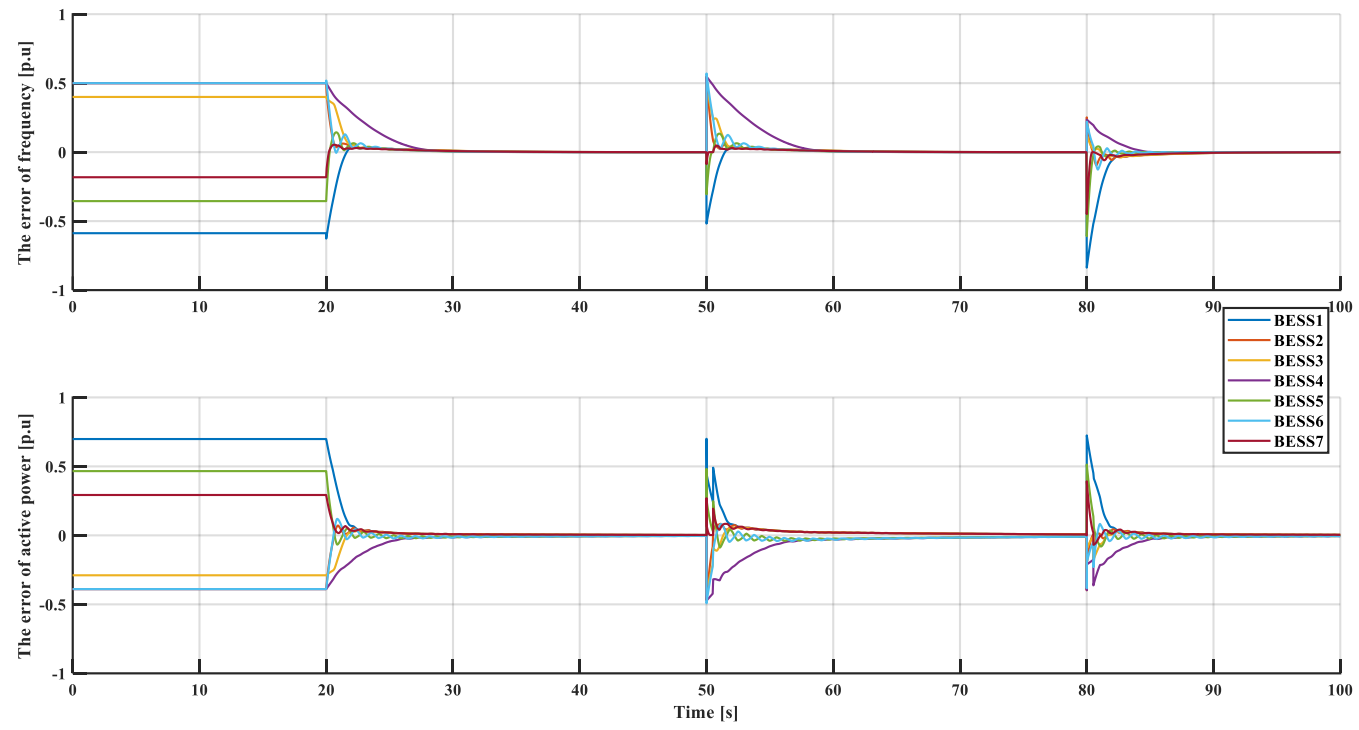

Fig. 13. The simulation results of the real-time errors of the frequency and active power sharing of each BESS under the communication delay in Case 3

## V. Conclusion

In this paper, all BESSs are classified by the CAK in consideration of the geographical factor. Accordingly, a distributed finite-time step-by-step consensus scheme is proposed and applied in the multi heterogeneous BESSs system based on the droop controller, of which the active/reactive power, the frequency, the voltage and the energy level of each BESS can reach consensus by only communicating with its neighbors' information in the finite time. Consequently, the active/reactive power and the energy level of each BESS convergence to the average values of their initial values respectively. And at the same time, the frequency and the voltage restoration to the nominal values is guaranteed, and which is not only the necessary but also crucial condition for the grid connection. Therefore, all BESSs can work in both the island and grid-connected modes, which makes our scheme flexible. In addition, the consensus inputs proposed is robust to the communication delay and the communication topology change, and this is critical in the engineering cases. Finally, these conclusions stated above are tested effectivley in the modified IEEE 57-bus system.

## Appendix

With **Lemmas 1** to **4** in hand, we can start the proof of **Theorem 1** now.

**Proof.** In the five inputs designed, there are actually only two forms in general. One is described in $u_i^Q$, $u_i^E$ and $u_i^P$, and the other is formulated in $u_i^{\omega}$ and $u_i^V$. Thus, $u_i^P$ and $u_i^{\omega}$ are taken to prove the convergence for simplicity.

a). Considering the real-time error $\Delta P_i(t)$ of active power, there is

$$\Delta P_i(t) = P_i(t) - \frac{1}{n}\sum_{i=1}^{n} P_i(t) \tag{21}$$

According to previous study in [36], active power can reach average consensus by implementing $u_i^P$. So, $\frac{1}{n}\sum_{i=1}^{n} P_i(t)$ is a constant value. Then, the differential of $\Delta P_i$ is shown as follow

$$\Delta \dot{P}_i(t) = \dot{P}_i(t) = u_i^P = -\sum_{j=1}^{n} w_{ij}\,\text{sgn}(P_i - P_j)\left|P_i - P_j\right|^{\eta} = -\sum_{j=1}^{n} w_{ij}\,\text{sgn}(\Delta P_i - \Delta P_j)\left|\Delta P_i - \Delta P_j\right|^{\eta} \quad (22)$$

Construct a Lyapunov function as shown (23)

$$V_P = K_P \Delta P^T \Delta P = K_P \sum_{i=1}^{n} \Delta P_i^2(t) \quad (23)$$

Where $\Delta P = [P_1(t), P_2(t), \cdots, P_n(t)]^T$ is the real-time error vector of active power sharing.

Thus, the derivative function of $V_P$ with respect to time is presented in (24)

$$\begin{aligned}\dot{V}_P &= 2K_P \sum_{i=1}^{n} \Delta P_i(t) \Delta \dot{P}_i(t) \\ &= -2K_P \sum_{i=1}^{n} \Delta P_i(t) \sum_{j=1}^{n} a_{ij}\,\text{sgn}(\Delta P_i - \Delta P_j)\left|\Delta P_i - \Delta P_j\right|^{\eta} \\ &= -K_P \sum_{i,j=1}^{n} \Delta P_i(t) a_{ij}\,\text{sgn}(\Delta P_i - \Delta P_j)\left|\Delta P_i - \Delta P_j\right|^{\eta} \\ &\quad - K_P \sum_{i,j=1}^{n} \Delta P_j(t) a_{ji}\,\text{sgn}(\Delta P_j - \Delta P_i)\left|\Delta P_j - \Delta P_i\right|^{\eta} \\ &= -K_P \sum_{i,j=1}^{n} (\Delta P_i - \Delta P_j) a_{ij}\,\text{sgn}(\Delta P_i - \Delta P_j)\left|\Delta P_i - \Delta P_j\right|^{\eta} \\ &= -K_P \sum_{i,j=1}^{n} a_{ij}\left|\Delta P_i - \Delta P_j\right|^{\eta+1}\end{aligned} \quad (24)$$

According to Lemma 1, if $1 < \eta + 1 < 2$, i.e. $0 < \eta < 1$, the inequality (25) can be established

$$\left(\sum_{i,j=1}^{n} \left(a_{ij}^{\frac{1}{\eta+1}} \left|\Delta P_i - \Delta P_j\right|\right)^{\eta+1}\right)^{\frac{1}{\eta+1}} \geq \left(\sum_{i,j=1}^{n} \left(a_{ij}^{\frac{1}{\eta+1}} \left|\Delta P_i - \Delta P_j\right|\right)^2\right)^{\frac{1}{2}} \quad (25)$$

Denote $A_P = (a_{ij}^{\frac{2}{\eta+1}})_{n\times n}$ an adjacency matrix, whose corresponding Laplacian matrix is $L_p$. Let $\lambda_P$ be the smallest positive eigenvalue of $L_p$ and set $K_P = 2\lambda_P$. From Lemma 2 and (25), we can obtain (26)

$$\begin{aligned}\dot{V}_P &\leq -2\lambda_P \left(\sum_{i,j=1}^{n} \left(a_{ij}^{\frac{2}{\eta+1}} \left|\Delta P_i - \Delta P_j\right|\right)^2\right)^{\frac{\eta+1}{2}} \\ &= -2\lambda_P (2\Delta P^T L_P \Delta P)^{\frac{\eta+1}{2}} \\ &\leq -2\lambda_P (2\lambda_P \Delta P^T \Delta P)^{\frac{\eta+1}{2}} = -2\lambda_P (V_P)^{\frac{\eta+1}{2}}\end{aligned} \quad (26)$$

By using Lemma 3 for $0 < \eta < 1$, i.e. $\frac{1}{2} < \frac{\eta+1}{2} < 1$, the active power can reach consensus in finite time $T_P$. And the convergence time $T_P$ can be calculated by (27)

$$T_P(\eta) = \frac{V_P^{\frac{1-\eta}{2}}(0)}{2\lambda_P \frac{1-\eta}{2}} = \frac{(2\lambda_P \left\|\Delta P(0)\right\|^2)^{\frac{1-\eta}{2}}}{(1-\eta)\lambda_P} \quad (27)$$

It is not so difficult to know that the convergence time is only depend on the initial errors and $\eta$ for a fixed communication topology.

That completes the first half proof of **Theorem 1**. Then we will come to the second half proof of **Theorem 1**.

b). Considering the real-time error $\Delta\omega_i(t)$ of frequency in (28), we can obtain (29)

$$\Delta\omega_i(t) = \omega_i(t) - \omega^r \quad (28)$$

$$\begin{aligned}\frac{\Delta\omega_i(t)}{dt} &= \frac{d\omega_i(t)}{dt} = u_i^{\omega} - u_i^P \\ &= -\sum_{j=1}^{n} a_{ij}\,\text{sgn}(\omega_i^0 - \omega_j^0)\left|\omega_i^0 - \omega_j^0\right|^{\eta} \\ &\quad - a_{0i}[\text{sgn}(\omega_i^0 - \omega^r)\left|\omega_i^0 - \omega^r\right|^{\eta} - P_i] \\ &= -\sum_{j=1}^{n} a_{ij}\,\text{sgn}(\Delta\omega_i - \Delta\omega\ )\left|\Delta\omega_i - \Delta\omega_j\right|^{\eta} \\ &\quad - a_{0i}[\text{sgn}(\Delta\omega_i)\left|\Delta\omega_j\right|^{\eta} - P_i]\end{aligned} \quad (29)$$

Construct a Lyapunov function as follow

$$V_{\omega} = K_{\omega} \Delta\omega^T(t) \Delta\omega(t) = K_{\omega} \sum_{i=1}^{n} \Delta\omega_i^2(t) \quad (30)$$

Where $\Delta\omega = [\omega_1(t), \omega_2(t), \cdots, \omega_n(t)]^T$ is the real-time error vector of the frequency.

Simultaneously, the derivative expression of function $V_{\omega}$ with respect to time can be derived as

$$\begin{aligned}\dot{V}_{\omega} &= 2K_{\omega} \sum_{i=1}^{n} \Delta\omega_i(t) \Delta\dot{\omega}_i(t) \\ &= 2K_{\omega} \sum_{i=1}^{n} \Delta\omega_i(t)\dot{\omega}_i(t) = 2K_{\omega} \sum_{i=1}^{n} \Delta\omega_i(t)(u_i^{\omega} - u_i^P) \\ &= -2K_{\omega} \sum_{i=1}^{n} \Delta\omega_i(t) \sum_{i=1}^{n} a_{ij}\,\text{sgn}(\Delta\omega_i - \Delta\omega_j)\left|\Delta\omega_i - \Delta\omega_j\right|^{\eta} \\ &\quad - 2K_{\omega} \sum_{i=1}^{n} \Delta\omega_i(t) a_{0i}\,\text{sgn}(\Delta\omega_i)\left|\Delta\omega_i\right|^{\eta} \\ &= -2K_{\omega} \sum_{i,j=1}^{n} \Delta\omega_i a_{ij}\,\text{sgn}(\Delta\omega_i - \Delta\omega_j)\left|\Delta\omega_i - \Delta\omega_j\right|^{\eta} \\ &\quad - 2K_{\omega} \sum_{i=1}^{n} a_{0i}\left|\Delta\omega_i\right|^{\eta+1} \\ &= -K_{\omega} \sum_{i,j=1}^{n} a_{ij}\left|\Delta\omega_i - \Delta\omega_j\right|^{\eta+1} - 2K_{\omega} \sum_{i=1}^{n} a_{0i}\left|\Delta\omega_i\right|^{\eta+1}\end{aligned} \quad (31)$$

In (31), the first half is the same as (24). For the second half, we have the diagonal matrix $A_0 = (a_{0ii}^{\frac{2}{\eta+1}})_{n\times n}$ and $a_{\omega ii} = a_{0i}$. Denote $\lambda_{\omega}$ the smallest positive eigenvalue of $(A_0 + L_p)$. Based on Lemma 1 and Lemma 4,

$$\begin{aligned}\dot{V}_{\omega} &\leq -K_{\omega}\left(\sum_{i,j=1}^{n} a_{ij}^{\frac{2}{\eta+1}} \left|\Delta\omega_i - \Delta\omega_j\right|^2 + 2\sum_{i=1}^{n} a_{ii}^{\frac{2}{\eta+1}} \left|\Delta\omega_i\right|^2\right)^{\frac{\eta+1}{2}} \\ &= -K_{\omega}[2\Delta\omega^T(A_0 + L_p)\Delta\omega]^{\frac{\eta+1}{2}} \\ &\leq -K_{\omega}[2\lambda_{\omega}\Delta\omega^T\Delta\omega]^{\frac{\eta+1}{2}} = -K_{\omega}[V_{\omega}]^{\frac{\eta+1}{2}}\end{aligned} \quad (32)$$

By using Lemma 3 for $0<\eta<1$ ,i.e. $\frac{1}{2}<\frac{\eta+1}{2}<1$ , the frequency can track the leader after a certain time $T_{\omega}$ , which can be calculated by (33)

$$T_{\omega}(\eta)=\frac{V_{\omega}^{\frac{1-\eta}{2}}(0)}{2\lambda_{\omega}\frac{1-\eta}{2}}=\frac{(2\lambda_{\omega}\left\|\Delta\omega(0)\right\|^{2})^{\frac{1-\eta}{2}}}{(1-\eta)\lambda_{\omega}} \qquad (33)$$

And by using Lemma 3 for $0<\eta<1$ ,i.e. $\frac{1}{2}<\frac{\eta+1}{2}<1$, the frequency can track the leader after a certain time $T_{\omega}$ . Obviously, the finite time $T_{\omega}$ is only determined by the frequency error along with $\eta$ in a connected communication graph. With formula expression of (33), we complete the second half proof of **Theorem 1**.

Thus, we complete the whole process of **Theorem 1**, now.

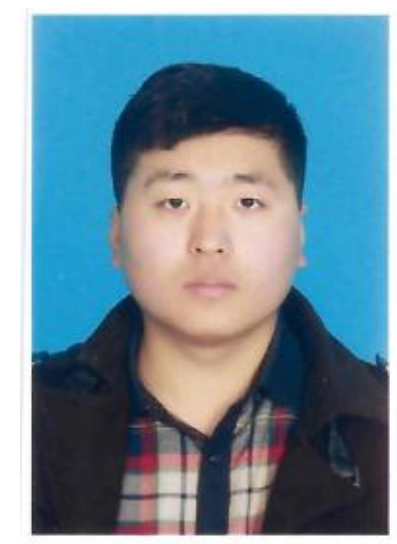

**Yalin Zhang** received a B.Sc. degree in Automation from Henan Polytechnic University, China, in 2016. He is now pursuing a M.Sc. degree in Control science and Engineering at Henan Polytechnic University, China. His research interests include modeling and control of complex system.

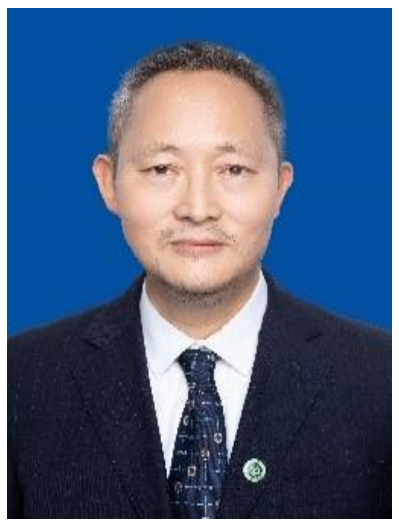

**Yunzhong Song** is now affiliated with the School of Electrical Engineering and Automation, Henan Polytechnic University as a full professor. He received a PhD degree from Zhejiang University, China in 2006. Professor Song has been a Visiting Professor in Chinese Academy of Sciences (CAS) and a Visiting Scholar in Umea University (Umeå Universitet), Sweden, and the winner of the 2017 Best Paper Award of International Conference on Artificial Life and Robotics (ICAROB) in Miyazaki, Japan. He has published almost 100 peer referenced journal and conference proceeding papers. His research interests include intelligent systems and networked systems control.

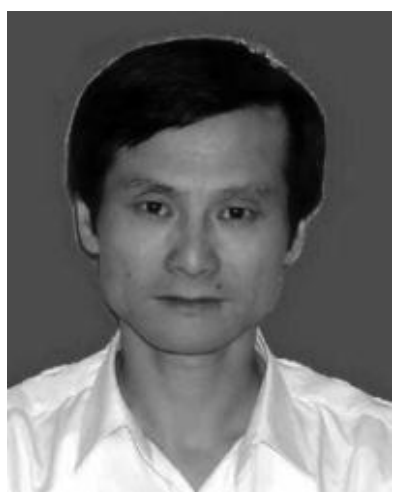

**Shumin Fei** received the Ph.D. degree from the Beijing University of Aeronautics and Astronautics, China, in 1995. From 1995 to 1997, he was a Post-Doctoral Research Fellow with Southeast University, where he is currently a Professor and PhD Advisor with the School of Automation，Southeast Uniersity. He has published more than 100 journal papers. His research interests include nonlinear systems, stability theory of delayed systems, and complex systems.